\documentclass{article}

\usepackage{graphicx}          

\usepackage[latin1]{inputenc} 
\usepackage{amssymb}          
\usepackage{amsmath}          
\usepackage{amsfonts}

\newtheorem{prosec}{Proposition}[section]   
\newtheorem{lemsec}{Lemma}[section]         
\newtheorem{corsec}{Corollary}[section]     

\usepackage[pdftex]{color}
\definecolor{Royalblue}{cmyk}{1,0.30,0.2,0.2}
\definecolor{oldgreen}{cmyk}{.55,0,0.5,0.4}
\newcommand{\agu}{\color{black}}
\newcommand{\mz}{\color{black}}
\newcommand{\ft}{\color{black}}

\newcommand{\proofn}{\noindent {\em Proof. }}

\newcommand{\Hi}{\mathcal{H}}

\newcommand{\C}{\mathbb{C}}

\newcommand{\R}{\mathbb{R}}
\newcommand{\Ec}{\mathcal{E}}

\newcommand{\Bi}{\mathcal{B}}

\newcommand{\Span}{\textrm{span}}

\newcommand{\tr}{\textrm{tr}}

\newcommand{\qed}{\hfill $\Box$ \vskip 2ex}

\newcommand{\tht}{\theta}

\newcommand{\beq}{\begin{equation}}
\newcommand{\eeq}{\end{equation}}
\newcommand{\beqa}{\begin{eqnarray}}
\newcommand{\eeqa}{\end{eqnarray}}
\newcommand{\beqan}{\begin{eqnarray*}}
\newcommand{\eeqan}{\end{eqnarray*}}
\newcommand{\bea}[1]{\begin{eqnarray}\label{#1}}
\newcommand{\eea}{\end{eqnarray}}

\newcommand{\be}[1]{\begin{equation}\label{#1}}
\newcommand{\ee}{\end{equation}}
\newcommand{\bp}[1]{\begin{prosec}\label{#1}}
\newcommand{\ep}{\end{prosec}}

\newcommand{\nn}{\nonumber}
\newcommand{\DD}{\mathfrak{D}(\Hi)}

\newcommand{\thtv}{\underline{\tht}}
\renewcommand\cal{\mathcal}

\newcommand{\intt}[1]         {\mathop{\rm\,int} \left( #1 \right)}         

\begin{document}

\title{On Quantum Channel Estimation\\ with Minimal Resources}
\author{Mattia Zorzi, Francesco Ticozzi, Augusto Ferrante \thanks{M. Zorzi, F. Ticozzi and A. Ferrante are  with the
Dipartimento di Ingegneria dell'Informazione, Universit\`a di
Padova, via Gradenigo 6/B, 35131 Padova, Italy {\tt\small
zorzimat@dei.unipd.it}, {\tt\small
ticozzi@dei.unipd.it}, {\tt\small
augusto@dei.unipd.it}}}
\date{}
\maketitle

\begin{abstract}
We determine the minimal experimental resources that ensure a
unique solution in the estimation of trace-preserving quantum
channels with both direct and convex optimization methods. A
convenient parametrization of the constrained set is used to
develop a globally converging Newton-type algorithm that ensures a
physically admissible solution to the problem. Numerical
simulations are provided to support the results, and indicate that
the minimal experimental setting is sufficient to guarantee good
estimates.
\end{abstract}


\section{Introduction}

{\agu Recent advances and miniaturization in laser technology and electronic
devices, together with some profound results in quantum physics and quantum information theory,
have generated in the last two decades increasing interest {\ft in the promising field of {\em quantum information engineering}. The potential of these new technologies have been demonstrated by a number of theoretical and experimental results, including intrinsically-secure quantum cryptography protocols, proof-of-principle implementation of quantum computing, as well as dramatic advances in controlled engineering of molecular dynamics, opto-mechanical devices, and many other experimentally available systems.
In this area, a key role is played by {\em control, estimation}  and {\em identification} problems for quantum-mechanical systems \cite{dalessandro-book,wiseman-book,petz-book,paris-book,holevo}. Important contributions to this multi-disciplinary research effort have been offered by a number control scientists, among which we would like to remember Mohammed Daleh.
A few examples dealing with control and estimation problems are
 \cite{Dahleh1996,Dahleh1996a,dalessandro-optimal,altafini-feedback,altafini-open,doyle-robust,vanhandel-feedback,nurdin-coherent,khaneja-timeoptimal,james-2008,ticozzi-markovian,belavkin-towards,Mabuchi2005}, and many more may be found in the surveys \cite{Khaneja-survey,Dong-Petersen-survey}.\\
In the spirit of developing research which is both strongly motivated by emerging applications and mathematically rigorous, we consider an identification problem arising in the}}
reconstruction of quantum dynamical models from experimental data. This is a key issue in many quantum information processing tasks \cite{nielsen-chuang,zeilinger,paris-book,lidar-tomography,boulant}.
For example, a precise knowledge of the behavior of a channel to be used for quantum computation or communications is needed in order to ensure that optimal encoding/decoding strategies are employed, and verify that the noise thresholds for hierarchical error-correction protocols, or for effectiveness of quantum key distribution protocols, are met \cite{nielsen-chuang,zeilinger}.
In many cases of interest, for example in communication in free-space \cite{villoresi}, channels are not stationary and to ensure good performances, repeated and fast estimation steps would be needed as a prerequisite for adaptive information encodings.
Motivated by these potential applications, we here focus on: (i) characterizing the minimum experimental setting needed for a consistent estimation of the channel;
(ii)  exploring how a minimal parametrization of the models can be exploited to reduce the complexity of the estimation algorithm; and (iii) testing (numerically) the {\em minimal experimental setting}, and compare it to ``richer'' experimental resources (in terms of available probe states and measured observables). {\ft In doing this, we present a general framework for the estimation of physically-admissible trace
preserving quantum channels by minimizing a suitable class of (convex) loss
functions which contains, as special cases, commonly used maximum likelihood (ML) functionals.}
In the large body of literature regarding channel estimation, or {\em quantum process tomography} (see e.g. \cite{paris-book,lidar-tomography} and references therein), the experimental resources are usually assumed to be given. {\ft Mohseni} et al. \cite{lidar-tomography} compare different strategies, but focus on the role of having entangled states as an additional resource, while we shall assume there is no additional quantum system to work with. {\ft The problem we study is closer in spirit to that taken in \cite{minimalstate} while studying minimal {\em state} tomography.}
\\
Our result include the determination of the minimal experimental resources (or {\em quorum}, in the language of \cite{paris-quorum}) for {\em Trace-Preserving (TP) channels estimation}, as part of a thorough theoretical analysis of both the inversion (direct, or standard tomography) method and a class of convex methods, including the widely-used maximum likelihood approach. The method we propose constrains the set of channels of the optimization problem to TP maps from the beginning, as opposed to the most common approach that introduces the TP constraint through a Lagrange multiplier \cite{paris-book,lidar-tomography}. This allow for an immediate reduction from $d^4$ to $d^4-d^2$ parameters in estimation problem. Our analysis can also be considered as complementary to the one presented in \cite{pino-tomography}, where the TP assumption is relaxed to include losses.
We provide a rigorous presentation of the results and we try, whenever possible, to make contact with ideas and methods of (classical) system identification. We next exploit the same convenient parametrization for TP channels we use in the theoretical analysis for developing a Newton-type algorithm with barriers, which ensures convergence in the set of physically-admissible maps. Numerical simulations are provided, confirming that the standard tomography method quite often fails to provide a positive map, and showing that experimental settings richer than the minimal one (in terms of input states and observables) do not lead to better performances (fixed the total number of available "trials").

\section{{\ft Quantum channels and $\chi$-representation}}


Consider a $d$-level quantum system  with associated Hilbert space $\Hi$ isomorphic to ${\mathbb C}^d$. The {\em state} of the system is described by a  density operator, namely by a positive, unit-trace matrix {\ft \[\rho\in\DD=\{\rho\in\C^{d\times d}| \rho=\rho^\dag\geq 0,\,\tr(\rho)=1\},\]} which plays the role of probability distribution in classical probability. A state is called {\em pure} if it has rank one, and hence it is represented by  an orthogonal projection matrix on a one-dimensional subspace. {\ft Measurable quantities or {\em observables} are associated with Hermitian matrices $X=\sum_kx_k\Pi_k,$ with $\{\Pi_k\}$ the associated spectral family of orthogonal projections. Their spectrum $\{x_k\}$ represents the possible outcomes, and the probability of observing the $k$th outcome can be computed as $p_\rho(\Pi_k)=\tr(\Pi_k\rho).$

A quantum channel (in Schr\"odinger's picture) is a map $
\Ec : \DD \rightarrow \DD.$ }It is well known \cite{kraus,nielsen-chuang} that a physically admissible quantum channel must be linear and {\em
Completely Positive} (CP), namely it admits an Operator-Sum Representation (OSR) \beq
\Ec(\rho)= \sum_{j=1}^{d^2}K_j\rho K_j^\dag\label{c1}\eeq where
$K_i\in \C^{d \times d}$ are called {\em Kraus operators}. In order to be
 {\em Trace Preserving} (TP), a necessary condition to map states to states, it must also hold that \beq \sum_{j=1}^{d^2}K_j^\dag
K_j=I_d \label{c2}\eeq
where $I_d$ is the $d\times d$ identity matrix.

An alternative way to describe a CPTP channel is offered by the
{\em $\chi$-representation}. Each Kraus operator $K_j\in
\C^{d\times d}$ can be expressed as a linear combination (with
complex coefficients) of $\{F_m\}_{m=1}^{d^2}$, $F_m$ being the
elementary matrix $E_{jk},$ with $m=(j-1)d+k.$ Accordingly, the
OSR (\ref{c1}) can be rewritten as \beq \label{chi1}
\Ec(\rho)=\sum_{m,n=1}^{d^2} \chi_{m,n}F_m\rho F_n^\dag \eeq where
$\chi$ is the $d^2 \times d^2$ Hermitian matrix with element
$\chi_{m,n}$ in position $(m,n)$. It easy to see that it must satisfy
\beq\chi=\chi^\dag\geq 0 \label{chi2}\eeq  and (following from
\eqref{c2}) \beq \sum_{m,n=1}^{d^2} \chi_{m,n}F_n^\dag
F_m=I_2\label{chi3}.\eeq  The map $\Ec$ is
completely determined by the matrix $\chi$.

{\ft We now introduce an helpful lemma which provides us with a
parametrization of trace preserving maps, and an easy formula for
computing probabilities in terms  $\chi$  \footnote{{\ft These
results implicitly relate the $\chi$ matrix emerging from the
basis of elementary matrices we chose to the Choi matrix $C_{\cal
E}=\sum_{mn}E_{mn}\otimes \Ec({E_{mn})}$\cite{petz-book}. In fact,
either by direct computation or by confronting formula
\eqref{lemma1} with its equivalent for the Choi matrix $C_\Ec$
(see e.g. \cite{paris-book}, chapter 2), it is easy to see that
$C_\Ec=O\chi O^\dag,$ where $O$ is the unique unitary such that
$O( X\otimes Y) O^\dag=Y\otimes X$ \cite{bhatia}.}}. For a brief
review of the partial trace definition and properties, see the
Appendix \ref{tr_parziale}.
\begin{lemsec}
\label{Chi}
{\agu
Let $\Ec_\chi$ be a CPTP map  associated with a given $\chi$.} Then for any $\rho\in\DD$
\beq\label{lemma1}\Ec_\chi(\rho)=\tr_2(\chi(I_d\otimes\rho^T)).\eeq
\end{lemsec}

\proofn Let us rewrite each $F_j$ as the corresponding elementary matrix $E_{lm},$ with $j=(l-1)d+m,\,k=(n-1)d+p,$ and relabel $\chi_{jk}$ as $\hat\chi_{lmnp}$ accordingly. Hence we get\beq\label{newchi}\chi=\sum_{l,m,n,p}\hat\chi_{lmnp}E_{ln}\otimes E_{mp},\eeq and
\[ \Ec_\chi(\rho)=\sum_{l,m,n,p} \hat\chi_{lmnp}E_{lm}\rho E_{pn}. \]
We can also expand $\rho=\sum_{rs}\rho_{rs}E_{rs},$ and substitute
it in the above expression. Taking into account that
$E_{lm}E_{np}=\delta_{mn}E_{lp}$, and defining
$[\hat\chi^B_{ln}]_{mp}=\hat\chi_{lmnp}$, we get:
\beqan\Ec_\chi(\rho)&=&\sum_{l,m,n,p,r,s}\rho_{rs}\hat\chi_{lmnp}E_{lm}E_{rs}E_{pn}\\&=&\sum_{l,n,r,s}\rho_{rs}\hat\chi_{lrns}E_{ln},\\
&=& \sum_{l,n}\left(\sum_{r,s}\rho_{rs}\hat\chi_{lrns}\right)E_{ln}\\
&=& \sum_{l,n}\left(\rho^T\hat\chi^B_{ln}\right)E_{ln},\\
&=& \tr_2(\chi(I\otimes\rho^T))\eeqan
where we used the fact that $\hat\chi^B_{ln}$ corresponds to the $d\times d$ dimensional block of $\chi$ in position $(l,n),$ and that for every pair of matrices $X,Y,$ we can write $\sum_{rs}X_{rs} Y_{rs}=\tr(X^TY).$ 
    \qed}

This leads to a useful expression for the computation of the expectations.
\begin{corsec}
Let us consider a state $\rho$, a projector $\Pi$ and a quantum channel $\Ec$
with associated {$\chi$-representation} matrix $\chi$. Then \[
{\ft p_{\Ec(\rho)}(\Pi)}=\tr(\Ec(\rho)\Pi)=\tr(\chi(\Pi\otimes\rho^T)).\]
\end{corsec}

\proofn It suffices to substitute \eqref{lemma1} in $p_{\chi,\rho}(\Pi)=\tr(\Ec(\rho)\Pi),$ and use the identity $\tr(X\otimes I) Y)=\tr(X\tr_2(Y)).$\qed

The TP condition \eqref{chi3} can also be re-expressed directly in terms of the $\chi$ matrix.

\begin{corsec}
\label{prop_Bjk} Let us consider a CP map $\Ec_\chi$ with associated {$\chi$-representation} matrix $\chi$. Then  $\Ec_\chi$ is TP if and only if
\beq\label{tpeq}\tr_1(\chi)=I_d.\eeq \end{corsec}

\proofn
Using the same notation we used in the proof of Lemma \ref{Chi}, we can re-espress the TP condition \eqref{chi3} as:
\[ I_d= \sum_{l,m,n,p}\hat\chi_{lmnp}E_{pn}E_{lm}
= \sum_{l,m,p}\hat\chi_{lmlp}E_{pm}
= \tr_1(\chi).\] \qed

\section{Identification Protocols}

Consider the following setting: a quantum system prepared in a {\em known pure state} $\rho$ is fed to an unknown channel ${\cal E}.$
The system in  the {\em output state}
$\Ec(\rho)$ is then subjected to a projective measurement of an {\em observable}: to our aim it will be sufficient to consider yes-no measurements associated to orthogonal projections $\Pi=\Pi^\dag=\Pi^2.$ Hence the outcome $x$ is in the set $\{0,1\},$ and can be
interpreted as a sample of the random variable $X$ which has {\ft distribution}  \beq P_{\chi(x),{\ft \rho}}=\left\{%
\begin{array}{ll}
p_{\chi,\rho}(\Pi) ,&\; \; \hbox{if } x=1\\
1-p_{\chi,\rho}(\Pi),&\; \; \hbox{if } x=0
\end{array}%
\right.\eeq where $p_{\chi,\rho}(\Pi)=\tr(\Ec_\chi(\rho)\Pi)$ is the probability that the
measurement of $\Pi$ returns outcome 1 when the state is $\Ec_\chi(\rho)$.

Assume that the experiment is repeated with a series of known
input (pure) states $\{\rho_k\}_{{ \mz k=1}}^{{\mz L}}$, and to
each trial the same orthogonal projections $\{\Pi_j\}_{{ \mz
j=1}}^{{\mz M}}$ are measured $N$ times, obtaining a series of
outcomes $\{x_{l}^{jk}\}$. We consider the sampled frequencies to
be our {\em data}, namely \beq f_{jk}:=\frac{1}{N}\sum_{l=1}^N
x_l^{jk}.\eeq
The channel identification problem (or as it is referred to in the physics literature, the {\em quantum process tomography} problem \cite{paris-book,nielsen-chuang,lidar-tomography}) we are concerned with
consists in constructing a {\em Kraus map} ${\Ec}_{\hat{\chi}}$ that fits the experimental data (in some optimal way), in particular estimating a matrix $\hat\chi$
satisfying constraints (\ref{chi1}),\eqref{chi2}.

\subsection{Necessary and sufficient conditions for identifiability}

It is well known \cite{sacchi-ML,paris-book} that by imposing
linear constraints associated to the TP condition \eqref{chi3}, or
equivalently \eqref{tpeq}, one reduces the $d^4$ real degrees of
freedom of $\chi$ to $d^4-d^2.$ This will be made explicit in the
following, by parameterizing $\chi$ in a ``generalized'' Pauli
basis (also known as gell-mann matrices, Fano basis or coherence
vector representation in the case of states
\cite{alicki-lendi,benenti-tomography,paris-book}). Usually the
trace preserving constraint is not directly included in the
standard tomography method \cite{lidar-tomography}, since in
principle it should emerge from the physical properties of the
channel, or it is imposed through a (nonlinear) Lagrange
multiplier in the maximum likelihood approach \cite{paris-book}.
Here, in order to investigate the minimum number of probe (input)
states and measured projectors needed to uniquely determine
$\chi$, it is convenient to include this constraint from the very
beginning. Doing so, we lose the possibility of exploiting a
Cholesky factorization in order to impose positive
semidefiniteness of $\chi$: noentheless, we show in Section
\ref{algorithm} that semidefiniteness of the solution can be
imposed algorithmically by using a barrier method {\mz
\cite{BOYD_VANDENBERGHE}}.

Consider an orthonormal basis for $d^2\times d^2$ Hermitian matrices of the form $\{\sigma_j\otimes\sigma_k\}_{j,k=0,1,\dots,d^2-1},$ where $\sigma_0=1/\sqrt{d}I_d,$ while $\{\sigma_j\}_{j=1,\dots,d^2-1}$ is a basis for the traceless subspaces. We can now write
\[\chi=\sum_{jk}s_{jk}\sigma_j\otimes\sigma_k.\]
If we now substitute it into \eqref{tpeq}, we get:
\[ I_d =\tr_1(\chi)=\sum_{jk}s_{jk}\tr(\sigma_j)\sigma_k \\
 =\sum_{k}\sqrt{d} \, s_{0k}\sigma_k,\]
and hence, since the $\sigma_j$ are linearly independent, we can conclude that $s_{00}=1,\,s_{0j}=0$ for $j=1,\ldots,d^2-1.$
Hence, the free parameters for a TP map (at this point not necessarily CP, since we have not imposed the positivity of $\chi$ yet) are $d^4-d^2,$ as we can write any TP $\chi$ as $\chi=d^{-1} I_{d^2}+\sum_{j=1,k=0}^{d^2{\mz -1},d^2{\mz -1}}s_{jk}\sigma_j\otimes\sigma_k,$ or, in a more compact notation,
\beq\label{repar} \chi(\underline \theta)=d^{-1} I_{d^2}+\sum_{\ell=1}^{d^4-d^2}\theta_\ell Q_\ell,\eeq
by rearranging the double index $j,k$ in a single $\ell,$ and defining the corresponding $Q_\ell=\sigma_j\otimes\sigma_k.$


The $\chi$ matrices corresponding to TP maps thus form an affine space. Let us call it linear part \[{\cal S}_{TP}=\Span\{\sigma_j\otimes\sigma_k\}_{j=1,\ldots ,d^2{\mz-1},k=0,\ldots,d^2{\mz -1}}.\]

It is convenient to define \beq\label{defB}B_{jk}=(\Pi_j-\frac{1}{d}I)\otimes\rho^T_k\eeq and ${\cal B}=\Span\{B_{jk} \}_{j=1,\ldots, {\mz M},k=1,\dots,{\mz L}}.$
Since we have $Q_\ell=\sigma_{j\neq 0}\otimes\sigma_k,$ it holds that \beq\label{zerotrace}\tr(Q_\ell (\Pi_j\otimes\rho^T_k))=\tr(Q_\ell B_{jk}).\eeq
Let us also introduce the function \beqan \hspace{2cm} g &:& \R^{d^4-d^2}
\rightarrow \R^{{\mz M\times L}}\nn\\
& {}& \thtv \mapsto g(\thtv)\eeqan with the component of
$g(\thtv)$ in position $(j,k)$ is defined as \beq
g_{jk}(\thtv)=p_{\chi(\thtv),\rho_k}(\Pi_j)=\tr(\chi(\thtv)(\Pi_j\otimes\rho^T_k)).\label{func_g}\eeq

\begin{prosec}\label{ginject} $g$ is injective if and only if $\mathcal{S}_{TP}\subset
\mathcal{B}$.\end{prosec} \proofn Given {\mz \eqref{func_g}}, we
have that
\beqan g_{jk}(\underline{\tht}_1)-g_{jk}(\underline{\tht}_2)&=& \tr[(\chi(\underline{\tht}_1)-\chi(\underline{\tht}_2))(\Pi_j\otimes\rho^T_k)]\\&=&\tr[S(\underline{\tht}_1-\underline{\tht}_2)B_{jk}]\nn\\
&=&\langle S(\underline{\tht}_1-\underline{\tht}_2),B_{jk}\rangle
\eeqan where $
S(\underline{\tht}_1-\underline{\tht}_2)=\chi(\underline{\tht}_1)-\chi(\underline{\tht}_2)=\sum_{l=1}^{d^4-d^2}(\tht_{1,l}-\tht_{2,l})Q_l\in
\mathcal{S}_{TP}.$ So, we have that \beq\label{eqinj}
g(\thtv_1)=g(\thtv_2) \; \Leftrightarrow \;  \langle
S(\underline{\tht}_1-\underline{\tht}_2),B_{jk}\rangle=0 \; \;
\forall \; j,k. \eeq Assume $\mathcal{S}_{TP}\subset
\mathcal{B}:$ the only element of ${\cal S}_{TP}$ for which the r.h.s. of \eqref{eqinj} could be true is zero. Since by definition $S(\underline{\tht}_1-\underline{\tht}_2)=0$ if and only if $\underline{\tht}_1=\underline{\tht}_2$ , $g$ is injective. On the other hand, assume that $\mathcal{S}_{TP}\not\subseteq
\mathcal{B}:$ therefore there exists $T \neq 0\in\mathcal{S}_{TP}\bigcap {\cal B}^\perp $ such that
\[ T=\sum_{\ell}\gamma_\ell Q_\ell,\quad
\langle T,B_{jk}\rangle=0 \; \forall j,k.\]
But this would mean that $\underline\tht$ and $\underline\tht+\underline\gamma$ have the same image $g(\underline\tht)$, and hence $g$ is not injective. \qed

This is a central result in our analysis: we anticipate here that
$g$ being injective is a necessary and sufficient condition for
{\em a priori} identifiability of $\chi,$ and thus for having a
unique solution of the problem for both inversion (standard
process tomography) and convex optimization-based (e.g. maximum
likelihood) methods, up to some basic assumptions on the cost
functional. The proof is given in full detail in Section
\ref{standard} and \ref{convex}.

As a consequence of these facts, we can determine the {\em minimal
experimental resources}, in terms of input states and measured
projectors, needed for faithfully reconstructing $\chi$ from
noiseless data $\{f^\circ_{jk}\}$, where
$f^\circ_{jk}=p_{\chi,\rho}(\Pi)$. In the light of proposition
\ref{ginject}, the minimal experimental setting is characterized
by a choice of $\{\Pi_j,\rho_k\}$ such that ${\cal S}_{TP}={\cal
B}$. Recalling the definition of ${\cal B},$ through \eqref{defB},
it is immediate to see that ${\cal S}_{TP}={\cal B}$ if and only
if $\Span\{\Pi_j-d^{-1}I_d\}=\Span\{\sigma_j,j=1,\ldots,d^2{\mz
-1}\}$ and $\Span\{\rho_k\}=\C^{d\times d}.$ We can summarize this
fact as a corollary of Proposition \ref{ginject}.
\begin{corsec} $g$ is injective if and only if we have {\em at least} $d^2$ linearly independent input states $\{\rho_k\},$ and $d^2-1$ measured $\{\Pi_j\}$ such that \[\Span\{\Pi_j-d^{-1}I_d\}=\Span\{\sigma_j,j=1,\ldots,d^2{\mz -1}\}.\]
\end{corsec}
We call such a set a {\em minimal experimental setting}. Notice that, using the terminology of \cite{paris-book,paris-quorum}, the minimal {\em quorum} of observables consists of $d^2-1$ properly chosen elements.  While in most of the literature at least $d^2$ observables are considered \cite{hradil-ML,lidar-tomography}, we showed it is in principle possible to spare a measurement channel at the output. A physically-inspired interpretation for this fact is that, since we {\em a priori} know, or assume, that the channel is TP, measuring the component of the observables along the identity does not provide useful information. This is clearly not true if one relaxes the TP condition, as it has been done in \cite{pino-tomography}: in that case, by the same line of reasoning, $d^2$ linearly independent observables are the necessary and sufficient for $g$ to be injective.

As an example relevant to many experimental situation, consider the qubit case, i.e. $d=2.$
A minimal set of projector has to span the traceless subspace of $\C^{2\times 2}$: one can choose e.g.:
\[\Pi_j=\frac{1}{2}I_2+\sigma_j,\;j=x,y,z.\]
\beq \label{4_3_nominale}
\rho_{x,y}=\frac{1}{2}I_2+\sigma_{x,y},\quad\rho_\pm=\frac{1}{2}I_2\pm\sigma_z.\eeq
It is clear that there is an asymmetry between the role of output and
inputs: in fact, exchanging the number of $\{\Pi_j\}$ and $\{\rho_k\}$ can
not lead to an injective $g$.

\subsection{Process Tomography by inversion}\label{standard}

Assume that $\mathcal{S}_{TP}\subset \mathcal{B},$ and that the
data $\{f_{jk}\}$ have been collected. { \mz Since $f_{jk}$ is an
estimate of $p_{\chi(\thtv),\rho_k}(\Pi_j)$, consider the
following least mean square problem \beq \min_{\thtv\in
\R^{d^4-d^2}} \|\underline{g}(\thtv)-\underline{f}\|\eeq where
$\underline{g}(\thtv)$ and $\underline{f}$ are the vectors
obtained by  stacking the $g_{jk}(\thtv)$ and $f_{jk}$
$j=1\ldots L,$ $k=\ldots M$, respectively. In view of (\ref{repar})
and (\ref{func_g}) we have that
$\underline{g}(\thtv)=T\thtv+d^{-1}\underline{1}$ where \beq
T=\left[\begin{array}{ccc}
\ddots & \vdots & \\
       & \tr(B_{jk}Q_\ell) & \\
       & \vdots & \ddots \end{array}\right] \eeq
and $\underline{1}$ is a vector of ones. {\ft Notice that the $\ell$th column of $T$ is formed with the inner products of $Q_\ell$ with each $B_{jk}.$ Since $\mathcal{S}_{TP}\subset
\Bi$, the $Q_\ell$ are linearly independent and the $B_{jk}$ are the generators of ${\cal B},$ then  $T$ is \em full column rank}}, namely has
rank $d^4-d^2$. Hence, in principle, one can reconstruct
$\hat\tht$ as
\beq\label{inversion}{\mz\hat\thtv}=T^\#(\underline{f}-\underline{1}),\eeq
$T^\#$ being the Moore-Penrose pseudo inverse of $T$
\cite{horn-johnson}. If the experimental setting is minimal, the
usual inverse suffices. However, as it is well known, when
computing $\chi$ this way from real (noisy) data, the positivity
character is typically lost \cite{paris-book,Aiello:06}. We better
illustrate this fact in Section \ref{numerics}, through numerical
simulations.

\subsection{Convex methods: general framework}\label{convex}

More robust approaches for the estimation of physically-acceptable
$\chi$ (or equivalent parametrizations) have been developed, most
notably by resorting to Maximum Likelihood methods
\cite{hradil-ML,sacchi-ML,paris-book,ziman-physicalmaps}. The
optimal channel estimation problem can be stated, by using the
parametrization for
$\chi(\underline\theta)=d^{-1}I_{d^2}+\sum_\ell\tht_\ell Q_\ell$
presented in {\mz the} previous section, as it follows: consider a
set of data $\{f_{jk}\}$ as above, and a cost functional
{\mz$J(\thtv):=h\circ g(\thtv)$ where $h: \R^{M\times
L}\rightarrow \R$ is a suitable function which depends on the data
$\{f_{jk}\}$.} We aim to find
\beq\label{optimal}{\mz\hat\theta=\arg\min_{\underline\theta}
J(\underline\tht)}\eeq subject to $\thtv$ belonging to some
constrained set ${\cal C}\subset \R^{d^4-d^2}$. In our case
\[{\cal C}={\cal A}_+\quad{\rm or }\quad{\cal C}={\cal A}_+ \cap {\cal I},\]
with ${\cal A}_+=\{{\mz\thtv}\;|\; \chi(\theta)\geq 0 \},$ while
${\cal I}=\{\thtv
\;|\;0<\tr(\chi(\underline\tht)(\Pi_j\otimes\rho^T_k))<1,\;
\forall\,j,k\}.$ The second constraint may be used when a cost
functional which is not well-defined for extremal probabilities, or in order to ensure that the estimated channel exhibits
some noise in each of the measured directions, as it is expected
in realistic experimental settings. Since the analysis does not
change significantly in the two settings, we will not distinguish
between them where it is not strictly necessary. The following
result will be instrumental to prove the existence of a unique
solution.

\begin{prosec}${\cal C}$ is a bounded set.
\end{prosec}
\proofn Since $\mathcal{C}\subset {\cal A}_+$, it is
sufficient to show that ${\cal A_+}$ is bounded or, equivalently, that a sequence
$\{\underline{\tht}_j\}_{j\geq 0}$, with $\underline{\tht}_j\in
\R^{d^4-d^2}$, and $\|\underline{\tht}_j\|\rightarrow +\infty$, cannot
belong to ${\cal A_+}$.
To this end, it is
sufficient to show that, as  $\|\underline{\tht}_j\|\rightarrow +\infty$, the minimum eigenvalue of
$\chi(\underline{\tht}_j)$ tends to $-\infty$ so that, for $j$
large enough, $\underline{\tht}_j$ does not satisfy condition
$\chi(\underline{\tht}_j)\geq 0$. {\agu
Notice that the map $\thtv \mapsto \chi(\thtv)$ is affine.
Moreover, since the $Q_\ell$s are lineraly independent, this map
is injective. Accordingly, $\|\chi(\underline{\tht}_j)\|$ approach
infinity as $\|\thtv_j\|\rightarrow +\infty$.} Since
$\chi(\underline{\tht}_j)$ is a Hermitian matrix,
$\chi(\underline{\tht}_j)$ has an eigenvalue $\lambda_j$ such that
$|\lambda_j|\rightarrow +\infty$ as
$\|\chi(\underline{\tht}_j)\|\rightarrow +\infty$. Recall that
$\chi(\underline{\tht}_j)$ satisfies (\ref{tpeq}) by construction
which implies that $\tr(\chi(\underline{\tht}_j))=d$ namely the
sum of its eigenvalues is always equal to $d$. Thus, there exists
an eigenvalue of $\chi(\underline{\tht}_j)$ which approaches
$-\infty$ as $j\rightarrow +\infty$, which is in contrast with its
positivity. So, {\mz$\mathcal{C}$} is bounded.\qed

Here we focus on the following issue: under which conditions on
the experimental setting (or, mathematically, on the set ${\cal
B}$ defined above) do the optimization approach have a unique
solution? In either of the cases above, ${\cal C}$ is the
intersection of convex nonempty sets: {\ft in fact, ${\cal S}_{TP}$ and
 $\chi\geq 0$ are convex and so must be the corresponding sets of $\thtv$, and it is immediate to verify that
${\cal I}$ is convex as well; all of these contain
$\underline\theta=0,$ corresponding to $\frac{1}{d}I_{d^2},$ and
hence they are non empty.} In the light of this, it is possible to
derive sufficient conditions on $J$ for existence and uniqueness
of the minimum in the presence of arbitrary constraint set ${\cal
C}.$ Define $\partial {\cal C}_0:=\partial {\cal C}\setminus
(\partial {\cal C}\cap {\cal A}_+)$
\begin{prosec}\label{unicita} Assume {\mz $h$ is continuous and strictly convex on $g({\cal C})$}, and
\beq \lim_{\underline{\tht}\rightarrow \partial \mathcal{C}_0
}J(\underline{\tht})=\lim_{\underline{\tht}\rightarrow \partial \mathcal{C}_0
}h\circ g(\underline{\tht})=+\infty. \label{cond2_per_unicita}\eeq If
$\mathcal{S}_{TP}\subset \mathcal{B}$, then the functional $J$ has
a unique minimum point in $\mathcal{C}$.\end{prosec} \proofn {\mz
Since $h$ is strictly convex on $g(\mathcal{C})$ and the linear function $g$, in view of
Proposition \ref{ginject}, is injective on
$\mathcal{C}$, $J$ is strictly convex on $\mathcal{C}$}. So, we
only need to show that $J$ takes a minimum value on $\mathcal{C}$.
In order to do so, it is sufficient to show that $J$ is
inf-compact, i.e., the image of $(-\infty,r]$ under the map
$J^{-1}$ is a compact set. Existence of the minimum for $J$ then
follows from a version of Weierstrass theorem since an inf-compact
function has closed level sets, and is therefore, lower
semicontinuous \cite[p. 56]{KOSMOL_OPTIMIERUNG}.
Define $\underline{\tht}_0:=\left(%
\begin{array}{ccccc}
  0 &  \ldots & 0\\
\end{array}%
\right)^T\in\R^{d^4-d^2}$. Observe that
$\chi(\underline{\tht}_0)={d}^{-1}I_{d^2}\geq 0$. Moreover, being
$\Pi_j\otimes\rho^T_k$ rank-one orthogonal projections \beq
\tr(\chi(\underline{\tht}_0)\Pi_j\otimes\rho^T_k)=\frac{1}{d} \;
\; \forall j,k.\eeq Therefore,
$\underline{\tht}_0\in{\mz\mathcal{C}}$ and call
$J(\underline{\tht}_0)=J_0<\infty$. So, we can restrict the search
for a minimum point to the image of $(-\infty,J_0 ]$ under
$J^{-1}$. Since ${\cal C}$ is a bounded set by construction, to
prove inf-compactness of $J$ it is sufficient to guarantee that
\beq\nonumber \lim_{\underline{\tht}\rightarrow
\partial \mathcal{C}_0}J(\underline{\tht})=+\infty.\eeq\qed

\subsection{Maximum Likelihood functionals}\label{ML_func}
{\mz \subsubsection{Binomial functional}} Assume a certain set of
data $\{f_{jk}\}$ have been obtained, by repeating $N$ times the
measurement of each pair $(\rho_k,\Pi_j)$. For technical reasons (strict convexity of the ML functional on the optimization set) and experimental considerations (noise typically irreversibly affects any state), it is typically assumed that $0<f_{jk}<1.$ The probability of
obtaining a series of outcomes with $c_{jk}=f_{jk}N$ ones for the
pair $(j,k)$ is then \beq P_\chi(c_{jk})= {N \choose c_{jk}}
\tr(\chi \Pi_j\otimes\rho_k^T)^{c_{jk}} [1-\tr(\chi
\Pi_j\otimes\rho_k^T)]^{N-c_{jk}} \label{p_chi} \eeq so that the
overall probability of {\mz $\{c_{jk}\}$}, may be expressed as:
$P_\chi({\mz \{c_{jk}\}})=\prod_{j=1}^{\mz L}\prod_{k=1}^{\mz M
}P_\chi(c_{jk}).$ By adopting the Maximum Likelihood (ML)
criterion, once fixed the $\{c_{jk}\}$ describing the recorded
data, the optimal estimate $\hat{\chi}$ of $\chi$ is given by
maximizing $P_\chi(\{c_{jk}\})$ with respect to $\chi$ over a
suitable set $\mathcal{C}$. Let us consider our parametrization of
the {TP} $\chi(\underline\tht)$ as in \eqref{repar} . If we assume
$0<\tr(\chi(\underline\tht) (\Pi_j\otimes\rho_k^T))<1,$ since the
logarithm function is monotone, it is equivalent (up to a constant
emerging from the binomial coefficients) to minimize over
$\mathcal{C}={\cal A}_+\cap{\cal I}$ \footnote{If the optimization is constrained to ${\cal A}_+\cap{\cal I},$ we are guaranteed that $f_{jk}$ will tend to be positive for a sufficiently large numbers of trials.} the function {\mz\beqa
J(\underline{\tht})&=&-\frac{1}{N}\log
P_{\chi(\thtv)}(\{c_{jk}\})+\sum_{j,k}\log\left(%
\begin{array}{c}
  N \\
  c_{jk} \\
\end{array}%
\right) \nn
\\&=&-\sum_{j,k} f_{jk}\log[\tr(\chi(\underline\tht)
(\Pi_j\otimes\rho_k^T)]\nn\\&&+(1-f_{jk})\log[1-\tr(\chi(\underline\tht)
(\Pi_j\otimes\rho_k^T))].\label{funz_Aiello}\eeqa}

{\mz Here, $h(X)=-\sum_{j,k}
f_{jk}\log(x_{jk})+(1-f_{jk})\log(1-x_{jk})$ with
$x_{jk}=[X]_{jk}$ and $X\in \R^{M\times L}$ is strictly convex on
$\R^{M \times L}$ because $0<f_{jk}<1$ by assumption.} Notice that
$\partial \mathcal{C}_0$ is the set of $\underline{\tht}\in {\cal
A}_+$ for which there exists at least one pair
$(\tilde{i},\tilde{k})$ such that
$\tr(\chi(\underline{\tht})(\Pi_{\tilde{j}}\otimes\rho_{\tilde{k}}^T))=0,1.$
Suppose that
$\tr(\chi(\underline{\tht})(\Pi_{\tilde{j}}\otimes\rho_{\tilde{k}}^T)\rightarrow0$
as $\underline{\tht}\rightarrow \partial \mathcal{C}_0$.
Therefore,
$\log[\tr(\chi(\underline{\tht})(\Pi_j\otimes\rho_k^T))]\rightarrow
-\infty$. Since $c_{\tilde{j},\tilde{k}}>0$ by assumption, we have
that {\mz\beqan \lim_{\underline{\tht}\rightarrow
  \partial \mathcal{C}_0} J(\underline{\tht})&=&-\hspace{-3mm}\lim_{\underline{\tht}\rightarrow
 \partial \mathcal{C}_0} \sum_{j,k}
f_{jk}\log[\tr(\chi(\underline{\tht})(\Pi_j\otimes\rho_k^T))]\\
&&+(1-f_{jk})\log[1-\tr(\chi(\underline{\tht}) (\Pi_j\otimes\rho_k^T))]\nn\\
&=&-f_{\tilde{j},\tilde{k}}
\hspace{-0mm}\lim_{\underline{\tht}\rightarrow
 \partial \mathcal{C}_0 }  \log[\tr(\chi(\underline{\tht})(\Pi_{\tilde{j}}\otimes\rho_{\tilde{k}}^T))]\\&=&+\infty.
 \eeqan}
In similar way, we obtain the same result from the other case, and the conditions for existence and uniqueness of the minimum of Proposition \ref{unicita}  are satisfied.

{\agu We now discuss {\em consistency} of this method. Let
$\thtv^\circ$ be the ``true" parameter and
$\chi=\chi(\thtv^\circ)$ be the corresponding $\chi$-matrix of the
``true''  channel. First observe that, once fixed the sample
frequencies  $f_{jk}$ (or, equivalently, $c_{jk}$),
$$J(\underline{\tht})\geq
-\sum_{j,k} f_{jk}\log[f_{jk}] +(1-f_{jk})\log[1-f_{jk}],$$ so
that if there exists $\hat \thtv\in{\cal C}$ such that
$\tr[\chi(\hat
\thtv)(\Pi_{\tilde{j}}\otimes\rho_{\tilde{k}}^T)]=f_{jk},$ then
such a $\hat \thtv$ is optimal. Hence, in particular, the (unique)
optimal solution corresponding to the $f_{jk}$ equal to the
``true" probabilities $\tr[\chi(\Pi_j\otimes\rho_k^T)]$ is exactly
$\thtv^\circ$. On the other hand, as the number of experiments $N$
increases, the sample frequencies  $f_{jk}$ tend to the ``true"
probabilities $\tr[\chi(\Pi_j\otimes\rho_k^T)]$. Therefore, in
view of convexity of $J$ and of the continuity of $J$ and its
first two derivatives, the corresponding optimal solution tends to
the ``true" parameter $\thtv^\circ$. This proves consistency.}


\subsubsection{Gaussian functional}
{\ft Assume a certain data $\{f_{jk}\}$ have been obtained. For each $\rho_k$ consider the sample vector
$\underline{f}_k=\left[%
\begin{array}{ccc}
 f_{1k} & \ldots & f_{Mk}\\
\end{array}%
\right]^T\in \R^{M}$, that can be
thought as a sample of $\underline{p}_{\chi}^k=\left[%
\begin{array}{ccc}
\tr(\chi(\Pi_1\otimes\rho_k^T)) & \ldots & \tr(\chi(\Pi_M\otimes\rho_k^T))\\
\end{array}%
\right]^T$. Accordingly, we can consider the probabilistic model
$\underline{f}_k=\underline{p}_{\chi}^k+\underline{v}_k$ where
$\underline{v}_k\sim \mathcal{N}(0,\Sigma),\Sigma>0$ is
gaussian noise.
This noise model is a good representation of certain experimental settings in quantum optics, where the sampled frequencies are obtained with high number of counts $c_{j}$ and the gaussian noise is due to the electronic of the measurement devices, typically photodiodes. In our model, we can think that to each measured $\Pi_j$ is associated a different device {\mz with} noise component $v_j$.} {\mz Notice that, the noise components are in general correlated.}
Let $\underline{\mathcal{D}}_j$ denote the device associated to
$\Pi_j$. Then, $\underline{\mathcal{D}}_j$ will measure the data
$f_{j1},\ldots,f_{jL}$. Since
$\underline{f}_{k}\sim\mathcal{N}(\underline{p}_\chi^k,\Sigma)$,
the probability of obtaining the outcomes $\underline{f}_k$ is
then \beq
P^k_\chi(\underline{f}_{k})=\frac{1}{\sqrt{(2\pi)^M\det\Sigma}}\exp\{-\frac{1}{2}(\underline{f}_{k}-\underline{p}_{\chi}^k)\Sigma^{-1}(\underline{f}_{k}-\underline{p}_{\chi}^k)^T\}
\eeq so that the overall probability of $\{f_{jk}\}$ is equal to
$P_\chi(\{f_{jk}\})=\prod_{k=1}^L P^k_\chi(\underline{f}_k)$. By
adopting the ML criterion, given $\{f_{jk}\}$, the optimal
estimate $\hat\chi$ of $\chi$ is given by maximizing
$P_\chi(\{f_{jk}\})$ with respect to $\chi$. Taking into account
the parametrization $\chi(\thtv)$ as in $(\ref{repar})$, it is
equivalent to minimize over $\mathcal{C}=\mathcal{A}_+$ the
function \begin{align}
J(\thtv)&=-2\log\left(\sqrt{(2\pi)^M\det(\Sigma)}\,
P_{\chi(\thtv)}(\{f_{jk}\})\right)\nn\\&=\sum_{k=1}^{L}(\underline{f}_{k}-\underline{p}_{\chi(\thtv)}^k)\Sigma^{-1}(\underline{f}_{k}-\underline{p}_{\chi(\thtv)}^k)^T.\end{align}
Then, it easy to see that the conditions of Proposition
\ref{unicita} are satisfied. Accordingly the minimum $\hat \thtv$
of $J$ is unique. Also in this case it is possible to show, along
the same lines used for the previous functional, the consistency
of the method.

\subsection{A convergent Newton-type algorithm}\label{algorithm}

In Section \ref{ML_func} we have presented two ML functionals and
showed the uniqueness of their solution. Now, we face the problem
of (numerically) finding the solution $\hat\thtv$ minimizing $J$
over the prescribed set. In
the following we will refer to the binomial functional
(\ref{funz_Aiello}), but the results can be easily extended for
the Gaussian case.

Consider $J$ as in (\ref{funz_Aiello}) and assume that
$\mathcal{S}_{TP}\subset \mathcal{B}$. Problem (\ref{optimal}),
with $\mathcal{C}=\mathcal{A}_+\cap\mathcal{I}$, is equivalent to
minimize $J$ over $\mathcal{I}$ with the linear inequality
constraint $\chi(\underline{\tht})\geq 0$. Rewrite the problem,
making the inequality constraint implicit in the objective
\beq\hat\thtv=\label{pb_vinc} \min_{\underline{\tht}\in
\mathcal{I}}J(\underline{\tht})+I_-(\underline{\tht})\eeq
where $I_-: \R^{d^4-d^2}\rightarrow \R$ is the indicator function for the non positive semidefinite matrices $\chi(\thtv)$ \beq I_-(\underline{\tht}):=\left\{%
\begin{array}{ll}
    0, & \hbox{$\underline{\tht}$ s.t. $\chi(\underline{\tht})\geq0$} \\
+\infty, & \hbox{elsewhere.} \\\end{array}%
\right.\eeq The basic idea is to approximate the indicator
function $I_-$ by the convex function \beq
\hat{I}_-(\underline{\tht}):=-\frac{1}{q}\log\det(\chi(\underline{\tht}))\eeq
where $q>0$ is a parameter that sets the accuracy of the
approximation (the approximation becomes more accurate as $q$
increases). Then, we take into account the approximated problem
\beq \hat{\underline{\tht}}^q=\min_{\thtv\in \intt{\mathcal{C}}}
G_q(\underline{\tht})\label{pb_barrier_approx}\eeq where
$\intt{\mathcal{C}}$ denotes the interior of $\mathcal{C}$ and the
convex function \beq G_q(\underline{\tht}):=q
J(\underline{\tht})-\log\det(\chi(\underline{\tht})). \eeq The
solution $\hat\thtv^q$ can be computed employing the following
Newton algorithm with backtracking stage:
\begin{enumerate}
\item Set the initial condition  $\thtv_0\in \intt{\mathcal{C}}$.
\item At
each iteration, compute the Newton step  \beq \Delta \underline{\tht}_l
 =-H_{\underline{\tht}_l}^{-1} \nabla G_{\underline{\tht}_l}\in\R^{d^4-d^2} \label{passonewton}\eeq where
\beqan
&& \left[\,\nabla G_{\underline{\tht}}\,\right]_s:= \frac{\partial{G_q(\underline{\tht})}}{\partial{\tht_{s}}}\nn\\
&& \hspace{0.5cm}=q \sum_{j,k} \left\{\frac{1-f_{jk}}{1-\tr[\chi(\underline{\tht})B_{jk}]} -  \frac{f_{jk}}{\tr[\chi(\underline{\tht})B_{jk}]}\right\}\times\\
&&\times\tr(Q_sB_{jk})-\tr[\chi(\underline{\tht})^{-1}Q_s] \\
&& \left[\,H_{\underline{\tht}}\,\right]_{r,s}:=\frac{\partial{G_q(\underline{\tht})}}{\partial{\tht_{r}\tht_{s}}}\nn \\
\hspace{.5mm}& =& q \sum_{j,k} \left\{\frac{1-f_{jk}}{[1-\tr(\chi(\underline{\tht})B_{jk})]^2} +  \frac{f_{jk}}{[\tr(\chi(\underline{\tht})B_{jk})]^2}\right\}\times\\
&&\times\tr(Q_rB_{jk})\tr(Q_sB_{jk})+\tr[\chi(\underline{\tht})^{-1}Q_r\chi(\underline{\tht})^{-1}Q_s]
\eeqan are the element in position $s$ of gradient
(understood as column vector) and the element in position $(r,s)$
of the Hessian of $G_q$ both computed at $\underline{\tht}$.
\item Set $t^0_l = 1$, and let $t^{p+1}_l=t^p_l/2$ until all
the following conditions hold: \beqan &&
0<\tr[\chi(\underline{\tht}_l + t^p_l \Delta \underline{\tht}_l)B_{jk}]<1 \; \; \forall \; j,k
 \\   && \chi(\thtv_l + t^p_l \Delta \underline{\tht}_l)\geq0\\
&& G_q( \underline{\tht}_l + t^p_l \Delta \underline{\tht}_l )<
G_q(\underline{\tht}_l)+\gamma
  t^p_l \nabla G_{\underline{\tht}_l}^T \Delta \underline{\tht}_l \label{backcond2}
  \eeqan where $\gamma$ is a real constant, $0<\gamma<\frac{1}{2}$.
\item  Set $\underline{\tht}_{l+1} = \underline{\tht}_l + t^p_l \Delta \underline{\tht}_l\in \intt{\mathcal{C}}$.
                        \item Repeat steps 2, 3 and 4 until the   condition  $\|\nabla G_{\underline{\tht}_l}\| < \epsilon$  is satisfied, where $\epsilon$ is a (small)
tolerance threshold, then set $\hat{\underline{\tht}}^q= \underline{\tht}_l$.
 \end{enumerate}
{\ft This algorithm converges globally: In the first stage, it
converges in a linear way, while in the last stage, it does converge quadratically.} The
proof of these facts is postponed to Appendix \ref{conv_Newton}.
Then, it is possible to show \cite[p. 597]{BOYD_VANDENBERGHE} that
\beq J(\hat{\thtv}) \leq J(\hat{\thtv}^q)\leq
J(\hat{\thtv})+\frac{d^2}{q}. \eeq Hence, $d^2/q$ is the accuracy
(with respect to $\hat \thtv$) of the solution $\hat{\thtv}^q$
found. This method, however, works well only setting a moderate
accuracy.

An extension of the previous procedure is given by the Barrier
method \cite[p. 569]{BOYD_VANDENBERGHE} which solves
(\ref{pb_vinc}) with a specified accuracy $\xi>0$:
\begin{enumerate}
\item    Set the initial conditions $q_0>0$ and $\underline{\tht}^{q_0}=\left[%
\begin{array}{ccc}
   0 &\ldots & 0 \\
\end{array}%
\right]^T\in \intt{\mathcal{C}}$.
\item Centering step: At the $k$-th iteration compute $\hat{\thtv}^{q_k}\in \intt{\mathcal{C}}$ by minimizing $G_{q_k}$ with starting
point $\hat\thtv^{q_{k-1}}$ using the Newton method previously
presented.
\item Set $q_{k+1}=\mu q_{k}$.
\item Repeat steps 2 and 3 until the   condition  $ \frac{d^2}{q_k}< \xi$  is satisfied, then set $\hat{\thtv}=
\hat{\thtv}^{q_k}$.
 \end{enumerate} So, at each iteration we
 compute $\hat{\thtv}^{q_k}$ starting from the previously computed
 point $\hat{\thtv}^{q_{k-1}}$, and then increase $q_k$ by a factor
 $\mu>1$. The choice of the value of the parameters $q_0$ and
 $\mu$ is discussed in \cite[p. 574]{BOYD_VANDENBERGHE}.
Since the Newton method used in the centering step globally
converges, the sequence $\{\hat{\thtv^{q_k}}\}_{k\geq 0}$
converges to the unique
 minimum point $\hat{\thtv}$ of $J$ with accuracy $\xi$. Moreover,
 the number of centering steps required to compute $\hat\thtv$ with
 accuracy $\xi$ starting with $q_0$ is equal to $\left \lceil \frac{\log (d^2/q_0)}{\log \mu}
 \right\rceil+1$, \cite[p. 601]{BOYD_VANDENBERGHE}.

\section{Simulation results}\label{numerics}
In this section we use the following notation: \begin{itemize}
    \item {\em IN method} denotes the process tomography by inversion
    of Section \ref{standard}.
    \item {\em ML method} denotes the ML method, using the functional (\ref{funz_Aiello}) of
    Section \ref{ML_func}. {\mz Here, the solution is computed using the Barrier method of Section \ref{algorithm}
with $\xi=10^{-5}$.}\end{itemize}
\subsection{Performance comparison}
\begin{figure}[htbp]
\centering
\includegraphics[trim=0cm 0cm 0cm 0cm, clip=true, width=0.8\textwidth]{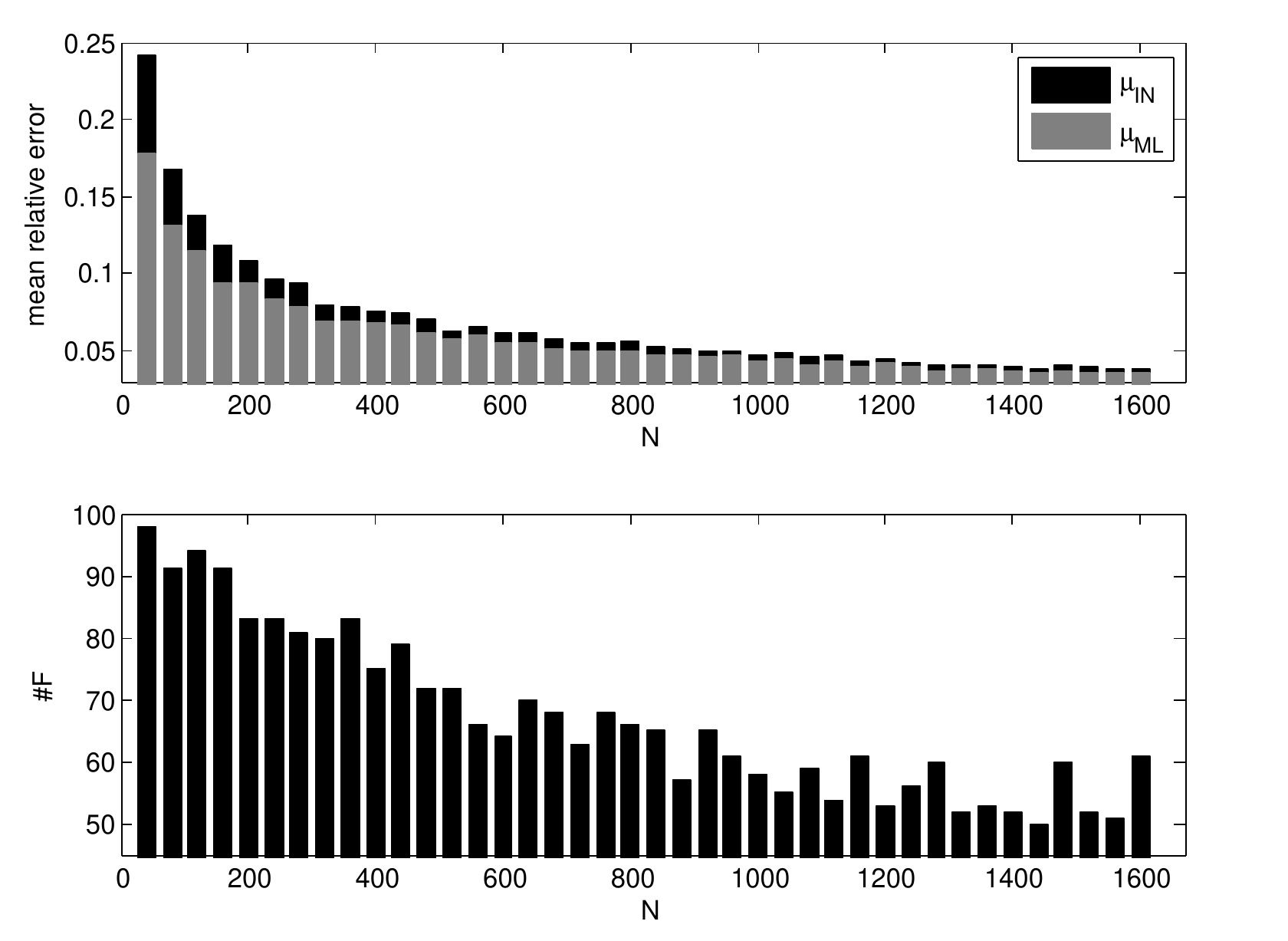} 
\vspace{-8mm} \caption{Comparison performance IN vs ML
method. $N$ is the total number of measurements {\mz for each $(\rho_k,\Pi_j)$}, $\mu$ is the mean relative error as introduced in \eqref{mu_norm}, while $\# F$ denotes the number of failures of the IN method, i.e. the times in which the reconstructed $\chi$ is not positive.}\label{figura1}
\end{figure}

Here, we want to compare the performance of IN and ML method for
the qubit case $d=2$. Consider a set of CPTP map
$\{\chi_l\}_{l=1}^{100}$ randomly generated and the minimal
setting (\ref{4_3_nominale}). Once the number of measurements $N$
for each couple $(\rho_k,\Pi_j)$ is fixed, we consider the
following comparison procedure:
\begin{itemize}
    \item At the $l$-th experiment, let $\{c_{jk}^l\}$ be the data corresponding to the map
    $\chi_l$. Then, compute the corresponding frequencies $f_{jk}^l=c^l_{jk}/N$
    \item From $\{f_{jk}^l\}$ compute the estimates
    $\hat\chi^{IN}_l$ and $\hat \chi^{ML}_l$ using IN and ML method
    respectively.
    \item Compute the relative errors  \beq e_{IN}(l)=\frac{\|\hat \chi^{IN}_l-\chi_l\|}{\|\chi_l\|},\;e_{ML}(l)=\frac{\|\hat \chi^{ML}_l-\chi_l\|}{\|\chi_l\|}. \label{e_norm}\eeq
    \item When the experiments are completed, compute the mean of
    the relative error \beq \mu_{IN}=\frac{1}{100}\sum_{l=1}^{100}
    e_{IN}(l),\;\mu_{ML}=\frac{1}{100}\sum_{l=1}^{100}
    e_{ML}(l).\label{mu_norm}\eeq
    \item Count the time that the IN method produces an estimate
    not positive semidefinite. This number is denoted as $\sharp
    F$.
\end{itemize} In Figure \ref{figura1}  is depicted the results obtained for different
lengths $N$ of measurements related to $\{c_{jk}^l\}$. The mean
error norm of ML method is smaller than the one corresponding to
the IN method, in particular when $N$ is small (typical situation
in the practice). In addition, more than half of the estimates
obtained by the IN method are not positive semidefinite, i.e not
physically acceptable, even when $N$ is sufficient large. Finally,
we observe that for both methods the mean error decrease as $N$
grows. This fact confirms in the practice their consistency.

\subsection{Minimal setting}
Let $\mathcal{T}_{M,L}$ denote the set of the experimental
settings with $L$ input states and $M$ observables satisfying
Proposition \ref{ginject}. Accordingly the set of the minimal
experimental settings is $\mathcal{T}_{d^2-1,d^2}$. Here, we
consider the case $d=2$. We want to compare the performance of the
minimal settings in $\mathcal{T}_{3,4}$ with those settings that
employ more input states and observables. We shall do so by
picking a test channel, finding a minimal setting that performs
well, and comparing its performance with a non minimal setting in
$\mathcal{T}_{M,L}$, $M>3,L\geq 4$ that performs well in this set
while the total number $N_T$ of trials is fixed.

\begin{figure}[htbp]
\centering
\includegraphics[trim=0cm 0cm 0cm 0cm, clip=true, width=0.8\textwidth]{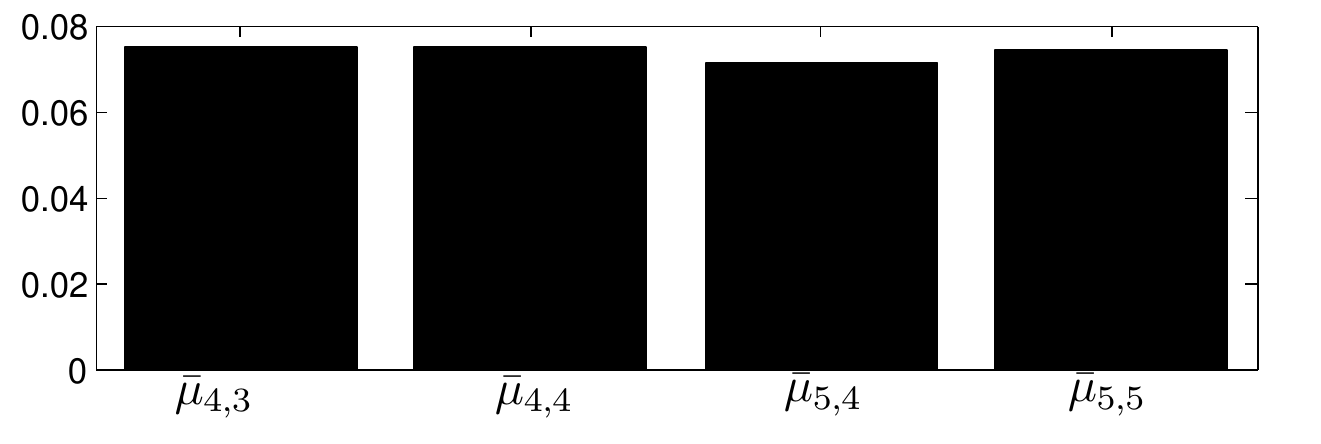} 
\vspace{-7mm} \caption{$\bar{\mu}_{M,L}$ for different values of
$M$ and $L$.}\label{figura3}
\end{figure}

\begin{figure}[htbp]
\centering
\includegraphics[trim=0cm 0cm 0cm 0cm, clip=true, width=0.8\textwidth]{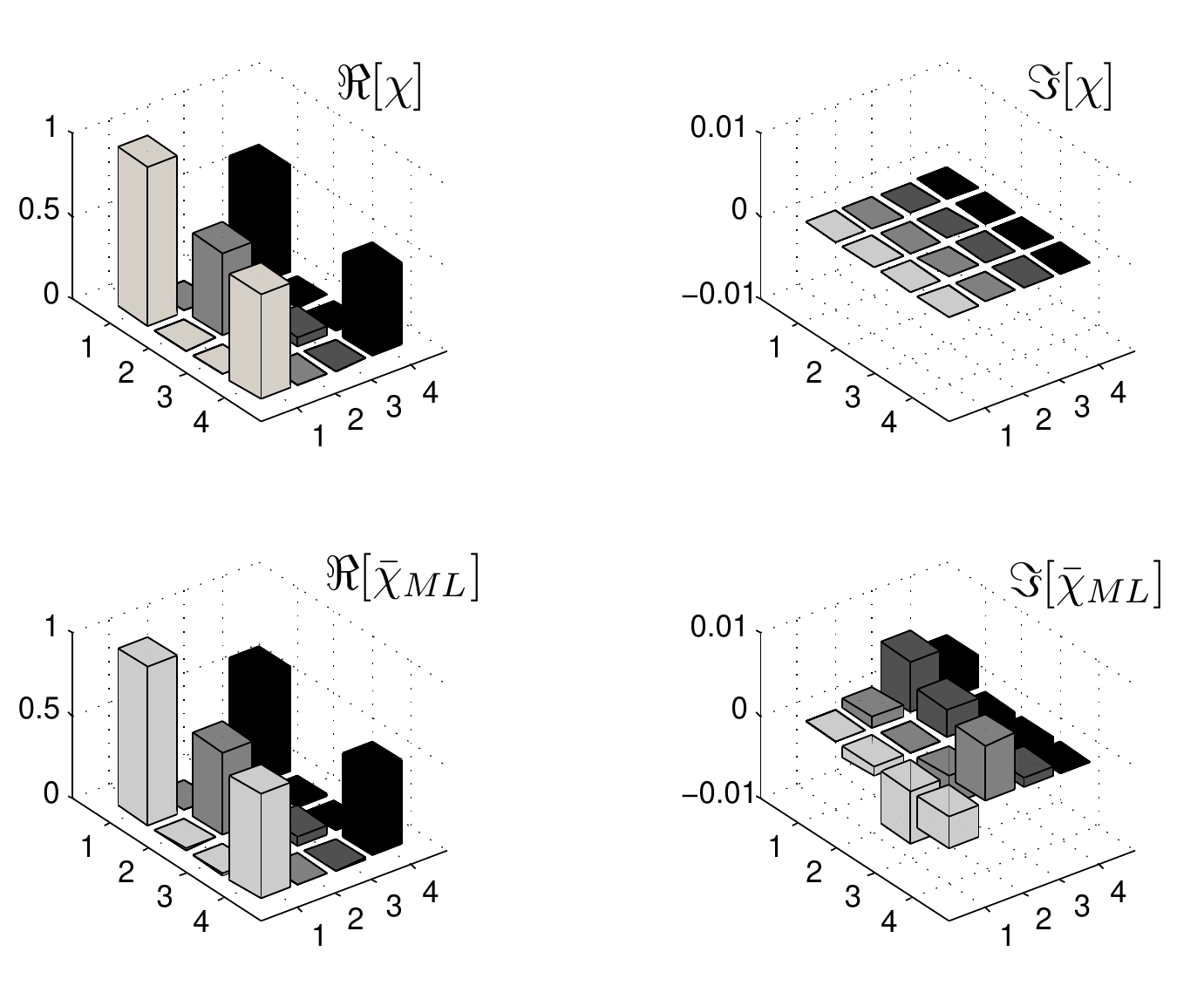} 
\vspace{-8mm} \caption{Real and imaginary part of $\chi$ (top) and the averaged estimation $\bar{\chi}_{ML}$ (bottom). {\ft In order to improve readability,} the vertical scale of the imaginary part has been magnified in order to show the errors are below 0.01.}\label{figura2}
\end{figure}

Consider the Kraus map (\ref{c1}) representing a perturbed
amplitude damping operation
($\gamma=0.5$) with \[ K_1=\sqrt{0.9}\left[%
\begin{array}{cc}
  \sqrt{0.5} & 0 \\
  0 & 0 \\
\end{array}%
\right], K_2=\sqrt{0.9}\left[%
\begin{array}{cc}
  1 & 0 \\
  0 & \sqrt{0.5} \\
\end{array}%
\right],  \]
$K_3=\sqrt{0.1}/2 I_2$, $K_j=\sqrt{0.1}/2\sigma_{l(j)}$, $j=4,5,6,$  $l(j)=x,y,z$ corresponding to the $\chi$-representation
\[ \chi=\left[%
\begin{array}{cccc}
0.95& 0  &  0 & 0.6364\\
  0 &  0.5&    0 &  0\\
 0 &        0  &  0.05&    0\\
    0.6364  &       0 &  0 &  0.5 \\
\end{array}%
\right].\]
We set the total number of trials $N_T=3600$. Fixed the set
$\mathcal{T}_{M,L}$ $M\geq 3$ $L\geq 4$, we take into account the
following procedure:
\begin{itemize}
\item Set $N=N_T\setminus (LM)$.
\item Choose a randomly generated collection $\{\mathrm{T}_m\}_{m=1}^{100}$, $\mathrm{T}_m\in \mathcal{T}_{M,L}$.
\item Perform $50$ experiments for each $\mathrm{T}_m$. At the $l$-th experiment we have a sample data $\{f_{jk}^m(l)\}$ corresponding to $\chi$ and $\mathrm{T}_m$. From $\{f_{jk}^m(l)\}$
compute the estimate $\hat\chi_{m}(l)$ using the ML method and the corresponding error norm $e_m(l)=\|\hat\chi_m(l)-\chi\| / \|\chi\|$.
\item When the experiments corresponding to $\mathrm{T}_m$ are completed, compute the mean error norm $\mu_m=\frac{1}{50}\sum_{l=1}^{50}e_m(l) $.
\item When we have $\mu_m$ for $m=1 \ldots100$, compute \[\bar{\mu}_{M,L}=\min_{m\in\{1,\ldots,100\}} \mu_m. \]
\end{itemize}
In Figure \ref{figura3},  $\bar{\mu}_{M,L}$ is depicted for different values of $M$ and $L$. As we can see, incrementing the number of input states/observables does not lead to an improvement in the performance index. Analogous results have been observed with other choices of test maps and $N_T$. Finally, in Figure \ref{figura2}  is depicted the true $\chi$ and the averaged estimation $\bar{\chi}_{ML}=\frac{1}{50}\sum_{l=1}^{50}\chi_m(l)$ with $m=\arg\min_{m\in\{1,\ldots,100\}} \mu_m$ for $M=3$ and $L=4$.\\


\section*{Acknowledgment}
The authors would like to thank Alberto Dall'Arche, Andrea
Tomaello, Prof. Paolo Villoresi and Dr. Giuseppe Vallone for
stimulating discussions on the topics of this paper. Work
partially supported by the QFuture research grant of the
University of Padova, and by the Department of Information
Engineering research project ``QUINTET''.

\appendix
\section{Partial trace}\label{tr_parziale}
 We here briefly recall the definition and some mathematical facts about the partial trace, without reference to its fundamental use in statistical quantum theory as the way to compute reduced (marginal) states, since we do not employ it to that scope. See e.g. \cite{nielsen-chuang,petz-book} for a comprehensive discussion.

Consider two finite-dimensional vector spaces ${\cal V}\,{\cal W},$ with $\dim{\cal V}=m,\,\dim{\cal W}=n.$
Let us denote by ${\cal M}_{j}$ the set of complex matrices of dimension $j\times j.$
Let $\{M_j\}$ be a basis for ${\cal M}_{m},$ and $\{N_j\}$ be a basis for ${\cal M}_{n},$ representing linear maps on ${\cal V}$ and ${\cal W},$ respectively.
Consider ${\cal M}_{mn}={\cal M}_{m}\otimes {\cal M}_{n}$: it is easy to show that the $m^2\times n^2$ linearly independent matrices $\{M_j\otimes N_k\}$ form a basis for ${\cal M}_{mn},$ where $\otimes$ denotes the Kronecker product.
Thus, one can express any $X\in{\cal M}_{mn}$ as \[X=\sum_{jk}c_{jk}M_j\otimes N_k.\]
The {\em partial trace} over ${\cal W}$ is the linear map
\beqan
\tr_{\cal W}&:&{\cal M}_{mn}\rightarrow {M}_m  \\
&&X\mapsto \tr_{\cal W}(X):= \sum_{j}(c_{jk}\tr(N_k))M_j.\eeqan
An analogous definition can be given for the partial trace over ${\cal V}.$
If the two vector spaces have the same dimension, $n=m$, we will indicate with $\tr_1$ and $\tr_2$ the partial traces over the first and the second spaces, respectively. The partial trace can be also implicitly defined (without reference to a specific basis) as the only linear function such that for any pair $X\in{\cal M}_{m},$  $Y\in{\cal M}_{n}$:
\[\tr_{\cal W}(X\otimes Y)=\tr(Y)X.\]
By linearity, this clearly implies \[\tr((A\otimes I)B)=\tr(A\,\tr_{2}(B)).\]


{\agu Notice that if $X\in{\cal M}_{mn}$, we may partition $X$ as an $m\times m$ block-matrix with block of size $n\times n$. In this way the partial trace with respect over the second space may be conveniently expressed as:
\beqan
\tr_{\cal W}(X)&=&\tr_{\cal W}\left[
\begin{array}{ccc}
X_{11}& \dots & X_{1m}\\
\vdots &\vdots&\vdots\\
X_{m1}& \dots & X_{mm}
\end{array}
\right]\\
&=&\left[
\begin{array}{ccc}
\tr(X_{11})& \dots & \tr(X_{1m})\\
\vdots &\vdots&\vdots\\
\tr(X_{m1})& \dots & \tr(X_{mm})
\end{array}
\right].
\eeqan
The partial trace with respect to $\cal V$,
$\tr_{\cal V}(X)$, is instead the $n\times n$ matrix having in position $j,k$
the trace of  the $m\times m$ matrix formed by selecting only the $(j,k)$ element of each of the blocks $X_{{\mz jk}}$.}

\section{Global convergence of the Newton algorithm}
 \label{conv_Newton} To prove the convergence of our Newton
algorithm we need of the following result.
 \begin{prosec}
Consider a function $f: X \subset \R^n\rightarrow \R$ twice
differentiable on $X$  with $H_x$ the Hessian of $f$ at $x$.
Suppose moreover that $f$ is strongly convex on a set $D\subset
X$,   i.e. there exists a constant $m> 0$ such that $H_x\geq mI$
for $x\in D $, and  $H_x$ is Lipschitz continuous on $ D $. Let
$\{x_i\}\in D$ be the sequence generated by the Newton algorithm.
Under these assumptions, Newton's algorithm with backtracking
converges globally. More specifically, $\{x_i\}$ decreases in
linear way for a finite number of steps, and converges in a
quadratic way to the minimum point after the linear stage.
\label{teoconv}
\end{prosec} \proofn See \cite[9.5.3, p. 488]{BOYD_VANDENBERGHE}. \qed
We proceed in the following way: Identify a compact set $D$ such
that $\underline{\tht}_l \in D$ and prove that the Hessian is
coercive and Lipschitz continuous on $D$. We then
apply Proposition \ref{teoconv} in order to prove the convergence.\\
Since $\thtv_0\in \intt{\mathcal{C}}$ we consider the set \beq
D:=\{\thtv\in \R^{d^4-d^2} \;|\; G_q(\thtv)\leq
G_q(\thtv_0)\}.\eeq The presence of the backtracking stage in the
algorithm guarantees that the sequence $G_q(\thtv_0),G_q(\thtv_1),
\ldots$ is decreasing. Thus $\thtv_l\in D$, $\forall l\geq 0$.
\begin{prosec}
 The following facts hold:
\begin{enumerate}
\item $D$ is a compact set.
  \item $H_{\thtv}$ is coercive and bounded on $D$, namely there exist $s,S>0$ such that \beq sI\leq
H_{\thtv} \leq SI \hspace*{1cm} \forall\; \thtv \in D. \eeq
\item $ H_{\thtv}$ is Lipschitz continuous on $D$.
\end{enumerate}
\end{prosec} \proofn 1) $D$ is contained into the
bounded set $\mathcal{C}$. Since $D$ is a finite dimensional
space, it is sufficient to show that \beq \lim_{\thtv\rightarrow
\partial \mathcal{C}} G_q(\thtv)=+\infty.\eeq
Here, we have three kind of boundary: $ \partial \mathcal{I}\cap
\intt{A_+}$, $\intt{\mathcal{I}}\cap \partial A_+$ and $
\partial \mathcal{I}\cap \partial A_+$. Notice that,  $\log\det(\chi(\thtv))$ takes finite
values on $\partial \mathcal{J}\cap \intt{ A_+}$. Accordingly,
taking (\ref{cond2_per_unicita}) into account, \beq
\lim_{\thtv\rightarrow
\partial \mathcal{I}\cap \intt{A_+}}G_q(\thtv)=q
\lim_{\thtv\rightarrow
\partial \mathcal{I}\cap \intt{A_+}} J(\thtv)=+\infty. \label{boundary2}\eeq
Then, $\intt{\mathcal{I}}\cap \partial A_+$ is the set of $\thtv $
for which $J$ is bounded and there exists at least one eigenvalue
of $\chi(\thtv)$ equal to zero. Thus, \beq \lim_{\thtv\rightarrow
 \intt{\mathcal{I}}\cap \partial
A_+}G_q(\thtv)=-\lim_{\thtv\rightarrow
 \intt{\mathcal{I}}\cap \partial
A_+}\log\det(\chi(\thtv))=+\infty.\label{boundary3}\eeq Finally,
from (\ref{boundary2}) and (\ref{boundary3}) it follows that
$G_q(\thtv)$ diverges as $\thtv$ approach $\partial
\mathcal{I}\cap
\partial A_+$.\\ 2) First, observe that $D\subset \intt{\mathcal{C}}$. Since $D$ is a
compact set, there exists $s>0$ such that \beq
\chi(\thtv)^{-1}\geq sI \;\; \forall \; \thtv\in D.  \eeq Define
\beqan &&
\delta_{jk}:=\frac{1-f_{jk}}{[1-\tr(\chi(\underline{\tht})B_{jk})]^2}
+
\frac{f_{jk}}{[\tr(\chi(\underline{\tht})B_{jk})]^2}>0\nn\\
&& [M_{jk}]_{r,s}:=\tr(Q_rB_{jk})\tr(Q_sB_{jk})\eeqan where
$M_{jk}$ is a positive semidefinite matrix with rank equal to one.
Accordingly, \beqan
[H_{\thtv}]_{r,s}&=& q \sum_{j,k}\delta_{jk}
[M_{jk}]_{r,s}+\tr[\chi(\underline{\tht})^{-\frac{1}{2}}Q_r\chi(\underline{\tht})^{-1}Q_s\chi(\underline{\tht})^{-\frac{1}{2}}]\nn\\
&\geq& q\sum_{j,k} \delta_{jk}[M_{jk}]_{r,s}
+s\tr[Q_r\chi(\underline{\tht})^{-1}Q_s]\\&\geq&
q\sum_{j,k}\delta_{jk}
[M_{jk}]_{r,s}+s^2\tr[Q_rQ_s]\nn\\
&\geq& q\sum_{j,k} \delta_{jk}[M_{jk}]_{r,s}+s^2\langle
Q_r,Q_s\rangle. \eeqan Since $\{Q_l\}_{l=1}^{12}$ are orthonormal
matrices and $\delta_{jk} M_{jk}\geq 0$, we have that \beq
H_{\thtv}\geq q\sum_{j,k} \delta_{jk}M_{jk}+s^2I \geq s^2 I.\eeq
Notice that, $H_{\thtv}$ is continuous on $\intt{\mathcal{C}}$. Since $D \subset \intt{\mathcal{C}}$, it
follows that $H_{\thtv}$ is continuous on the compact $D$. Hence,
there exists $S>0$ such that $H_{\thtv} \leq SI$ $\forall \;
\thtv\in D$. We
conclude that $ H_{\thtv}$ is coercive and bounded on $D$.\\
3) $H_{\thtv}$ is continuous on $D$ and $ \| H_{\thtv} \|\leq S $
$\forall \; \thtv\in D$, therefore $H_{\thtv}$ is Lipschitz
continuous on $D$. \qed Since all the hypothesis of the
Proposition \ref{teoconv} are satisfied, we have the following
proposition. \begin{prosec} The sequence $\{\thtv_l\}_{l\geq 0}$
generated by the Newton algorithm of Section \ref{algorithm}
converges to the unique minimum point
$\hat{\thtv}^q\in\intt{\mathcal{C}}$ of $G_q$.\end{prosec}

\bibliographystyle{plain}
\bibliography{biblio}

\end{document}